# 3D Visual Tracking to Quantify Physical Contact Interactions in Human-to-Human Touch

**Shan Xu[1], Chang Xu[1], Sarah McIntyre[2], Håkan Olausson[2], Gregory J. Gerling[1*]**

[1]School of Engineering and Applied Science, University of Virginia, Charlottesville, Virginia, USA

[2]Center for Social and Affective Neuroscience (CSAN), Linköping University, Sweden

**\* Correspondence:**
Gregory J. Gerling
gg7h@virginia.edu

## Abstract

Across a plethora of social situations, we touch others in natural and intuitive ways to share thoughts and emotions, such as tapping to get one's attention or caressing to soothe one's anxiety. A deeper understanding of these human-to-human interactions will require, in part, the precise measurement of skin-to-skin physical contact. Among prior efforts, each measurement approach exhibits certain constraints, e.g., motion trackers do not capture the precise shape of skin surfaces, while pressure sensors impede skin-to-skin contact. In contrast, this work develops an interference-free 3D visual tracking system using a depth camera to measure the contact attributes between the bare hand of a toucher and the forearm of a receiver. The toucher's hand is tracked as a posed and positioned mesh by fitting a hand model to detected 3D hand joints, whereas a receiver's forearm is extracted as a 3D surface updated upon repeated skin contact. Based on a contact model involving point clouds, the spatiotemporal changes of hand-to-forearm contact are decomposed as six, high-resolution, time-series contact attributes, i.e., contact area, indentation depth, absolute velocity, and three orthogonal velocity components, together with contact duration. To examine the system's capabilities and limitations, two types of experiments were performed. First, to evaluate its ability to discern human touches, one person delivered cued social messages, e.g., happiness, anger, sympathy, to another person using their preferred gestures. The results indicated that messages and gestures, as well as the identities of the touchers, were readily discerned from their contact attributes. Second, the system's spatiotemporal accuracy was validated against measurements from independent devices, including an electromagnetic motion tracker, sensorized pressure mat, and laser displacement sensor. While validated here in the context of social communication, this system is extendable to human touch interactions such as maternal care of infants and massage therapy.

## 1    Introduction

Social and emotional communication by touch is important to human development in daily life. It contributes to brain and cognitive development in infancy and childhood (Cascio et al., 2019), and plays a role in providing emotional support (Coan et al., 2006), and forming social bonds (Vallbo et al., 2016). For example, being touched by one's partner mitigates one's reactivity to psychological pressure, as observed in decreased blood pressure, heart rate, and cortisol levels (Gallace and Spence, 2010). Behaviors such as compliance, volunteering, and eating habits are also positively improved

(Gallace and Spence, 2010). Moreover, several works now indicate that particular social messages and emotional sentiments can be readily recognized from touch alone (Hertenstein et al., 2006, 2009; Thompson and Hampton, 2011; Hauser et al., 2019a; McIntyre et al., 2021). Despite their importance and ubiquity, we have just begun to quantify the exact nuances in the underlying physical contact interactions used to communicate affective touch.

To decompose how physical contact interactions evoke sensory and behavioral responses, most prior studies employ highly controlled stimuli, which vary a single factor at a time. In particular, mechanical and thermal interactions are typically delivered to a person's skin using robotically driven actuators (Löken et al., 2009; Essick et al., 2010; Ackerley et al., 2014a; Tsalamlal et al., 2014; Bucci et al., 2017; Teyssier et al., 2020; Zheng et al., 2020). For example, brush stimuli swept along an arc have been widely adopted to mimic caress-like stroking, while controlling their velocity, force, surface material, and/or temperature. Using such stimuli, C-tactile afferents are shown to be preferentially activated at stroke velocities around 1-10 cm/s, which align with ratings of pleasantness (Löken et al., 2009; Essick et al., 2010; Ackerley et al., 2014a). Beyond experiments to examine brush stroke, more complex interactions have been delivered via humanoid robots and robot hands (Teyssier et al., 2020; Zheng et al., 2020). However, device-delivered stimuli do not fully express the natural and subtle complexities inherent in human-to-human touch. This can result in disconnect with the everyday, real-world interactions for which our sensory systems are finely tuned.

Measuring and quantifying free and unconstrained human-to-human touch interactions is complex and challenging. In particular, the physical interactions are unscripted, unconstrained, and individualized with rapid and irregular transitions. Indeed, multiple contact attributes often co-vary over time, e.g., lateral velocity, contact area, indentation depth. Therefore, in moving toward quantification, the initial efforts used qualitative, manual annotation to describe touch gestures, and their contact intensity and duration (Hertenstein et al., 2006, 2009; Yohanan and MacLean, 2012; Andreasson et al., 2018). While adaptable to a wide range of touch interactions and settings, qualitative methods are constrained by the time required to analyze the data, the potential subjectivity of human coders, and a courser set of metrics and classification levels. For instance, contact intensity is typically classified in only three levels as light, medium, strong. As a result, automated techniques have been introduced, such as electromagnetic motion trackers (Hauser et al., 2019a; Lo et al., 2021) and sensorized pressure mats (Silvera-Tawil et al., 2014; Jung et al., 2015), with each their own capabilities and limitations. For instance, electromagnetic trackers capture the movement of only a handful of points, thus unable to monitor complex surface geometry, and can emit electromagnetic noise incompatible with sensitive biopotential recording equipment. Pressure sensors and mats inhibit direct skin-to-skin contact, when even thin films are shown to attenuate touch pleasantness (Rezaei et al., 2021). Three-dimensional optical tracking methods have also been employed, such as infrared stereo techniques (Hauser et al., 2019a, 2019b; McIntyre et al., 2021), motion capture systems (Suresh et al., 2020), and stereo cameras with DeepLabCut (Nath et al., 2019). While these methods are specialized in tracking joint positions of hands and limbs, they do not capture the shape and geometry of body parts, since the infrared cameras lack sufficient accuracy on depth, motion capture systems only track pre-attached markers, and stereo matching of multiple cameras often fail with texture-less surfaces. In contrast, depth cameras can provide high spatial resolution point clouds and allow shape extraction of texture-less body parts, such as a forearm. Depth cameras, as well, are more readily set up without calibration, afford minimum magnetic interference, and can be located at a larger distance from the area of interest. While depth cameras have been used in hand tracking and 3D reconstruction (Rusu and Cousins, 2011; Taylor et al., 2016), they have not been used to measure contact interactions in human-to-human touch.

While defined to a degree, we are still deciphering those physical contact attributes vital to social touch communication. In such settings, human touch interactions tend to include gesture, pressure/depth, velocity, acceleration, location, frequency, area, and duration (Hertenstein, 2002; Hertenstein et al., 2006, 2009; Yohanan and MacLean, 2012; Silvera-Tawil et al., 2014; Jung et al., 2015; Andreasson et al., 2018; Hauser et al., 2019a, 2019b; Lo et al., 2021; McIntyre et al., 2021). To understand the functional importance of specific movement patterns, certain attributes such as spatial hand velocity have been further decomposed into directions of normal and tangential (Hauser et al., 2019a) or forward-backward and left-right (Lo et al., 2021). Moreover, simultaneous tracking of multiple contact attributes is needed for understanding naturalistic, time-dependent neural output of peripheral afferents. For example, a larger contact area should recruit more afferents, larger force or indentation should generate higher firing frequencies, and optimal velocity in tangential direction should evoke firing of C-tactile afferents (Johnson, 2001; Löken et al., 2009; Hauser et al., 2019b).

Herein, we develop an interference-free 3D visual tracking system to quantify spatiotemporal changes in skin-to-skin contact during human-to-human social touch communication. Human-subjects experiments evaluate its ability to discern unique combinations of contact attributes used to convey distinct social touch messages and gestures, as well as the identities of the touchers. Moreover, the system's spatiotemporal accuracy is validated against measurements from independent devices, including an electromagnetic motion tracker, sensorized pressure mat, and laser displacement sensor.

## 2  Human-to-Human Contact Tracking System

This work introduces a 3D visual tracking system and data processing pipeline, which used a high-resolution depth camera to quantify contact attributes between the bare hand of a toucher and the forearm of a receiver. As illustrated in Figure 1, the tracking system captured the 3D shape and movements of the toucher's hand and the receiver's forearm independently but simultaneously within the same camera coordinate system. Physical skin contact was detected between the hand and forearm based on interactions of their 3D point clouds. Seven contact attributes were derived over the time course of touch, which were contact area, indentation depth, contact duration, overall contact velocity, and its three orthogonal velocity components.

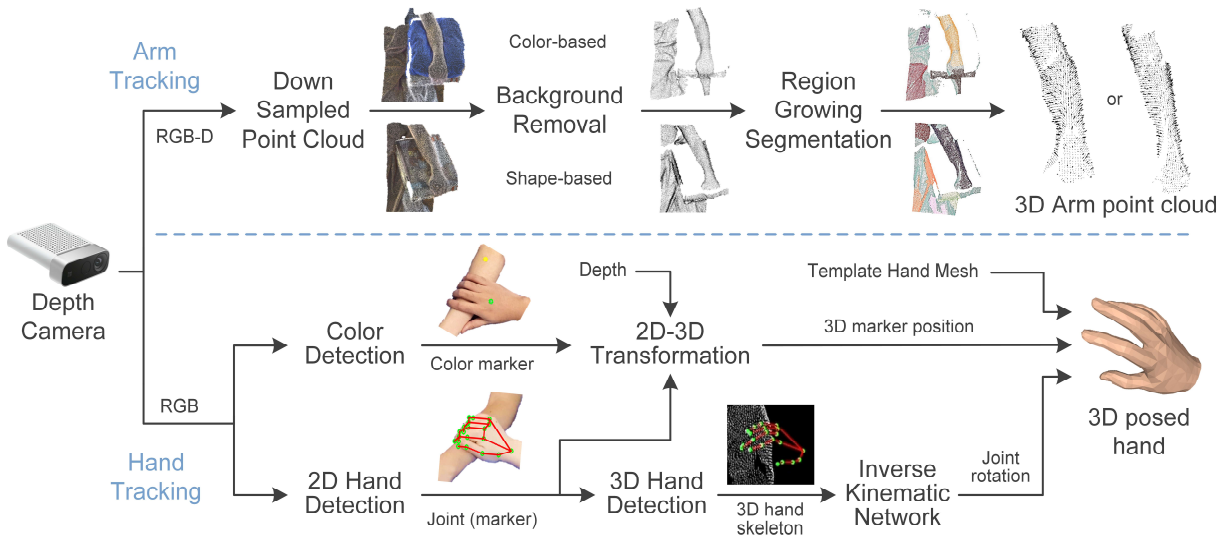

**Figure 1.** 3D visual tracking setup and data workflow. The toucher's hand and receiver's forearm are tracked using one depth camera (Microsoft Azure Kinect). Forearm shape is extracted as a point cloud while the hand mesh is animated by the gestures and movements of the toucher's hand.

## 2.1 3D Shape and Motion Tracking with Depth Camera

The tracking procedure extracts the detailed 3D shape of the touch receiver's forearm. By merging the camera's RGB and depth information, an RGB-D image was derived and then converted into a dense point cloud per frame. The point cloud was cropped and downsampled to balance information and computation costs. To obtain a clean point cloud of the forearm without background, neighboring points around the forearm were first removed. Two removal methods were used alternatively based on the experimental setup (Figure 1). If the receiver's forearm was placed on a flat surface, such as a table, the points within that flat surface could be removed in a shape-based manner using the plane model segmentation algorithm provided by the Point Cloud Library (PCL) (Rusu and Cousins, 2011). In the second case, if a monochromatic holder was set underneath the forearm, such as a cushion, then the points of that holder could be removed by color-based segmentation in the HSV color space. Next, the 3D region growing segmentation algorithm (Rusu and Cousins, 2011) was applied to separate the rest point cloud into multiple clusters according to the smoothness and distance between points. Since neighboring points around the forearm were removed in advance, points farther away in the background were assigned to separate clusters instead of being blended with the arm. Finally, by setting a relatively large smoothness threshold, all arm points could be grouped into one cluster despite the curvature of the forearm shape.

In human-to-human touch scenarios, the receiver's forearm is frequently occluded by the toucher's hand. Given that a blocked arm region is nearly impossible to capture, only the shape of the forearm prior to the contact was extracted. More specifically, the forearm point cloud was extracted before the beginning of each contact interaction to update its shape and position. During the contact, its position was refreshed in real-time according to the 3D position of the color marker on the arm, though its shape was not updated during the contact. Once the forearm was shape updated, the normal vector $\boldsymbol{n}_{arm}^i$ of each arm point $\boldsymbol{p}_{arm}^i$ was calculated and updated as well to facilitate further contact detection and measurement.

The hand tracking procedure was developed to capture the posture and position of the toucher's hand by combining depth information with a monocular hand motion tracking algorithm (Zhou et al., 2020). The algorithm is robust to occlusions and object interactions, which is advantageous in hand-arm contact. The monocular tracking algorithm contains two neural network modules to predict the 3D location and rotation of all 21 hand joints. In the first module of the hand joint detection network, features extracted from the 2D RGB image were first fed into a 2-layer convolutional neural network (CNN) to detect the probability of the 2D position of all joints. Then, another two 2-layer CNN was used to predict the 3D position of hand joints based on 2D features and 2D joint position estimates. In the second module of the inverse kinematic network, a 7-layer fully connected neural network was designed to derive the 3D rotation of each joint. Finally, the parametric MANO hand model (Romero et al., 2017) was employed to incorporate 3D joint rotations to animate the hand mesh following the shape and pose of the toucher's hand.

The rendered hand mesh was expressed in the local hand coordinate without the spatial information of the hand position. Therefore, depth information is incorporated here to locate the hand mesh in the camera coordinate, according to the movement of any hand joint or the color marker on the back of the hand (Figure 1). Specifically, the 2D position of the color marker was detected in the in the HSV, while the 2D position of the joint was retrieved from the detected 2D hand. The depth value of the hand joint or marker was derived by transforming the depth image to the RGB coordinate, which was then used to obtain its 3D position following the camera projection model. By identifying the corresponding point of that marker or joint in the hand mesh model, the posed hand mesh was moved in real-time following the toucher's hand movements.

## 2.2 Definition of Contact Attributes

Hand-arm contact was measured in a point-based manner, which afforded higher resolution compared with a geometry-based method (Hauser et al., 2019a). First, a contact interaction between the hand and forearm was detected when at least one vertex point of the hand mesh was underneath the arm surface. More specifically, for each hand vertex point $p^i_{hand}$, its nearest arm point $p^i_{arm}$ was found first. Then, as detailed in Equation (1), if the angle between the vector $p^i_{hand} - p^i_{arm}$ and the normal vector $n^i_{arm}$ of arm point $p^i_{arm}$ is larger than or equal to 90 degrees, this hand vertex is marked as underneath the arm surface.

$$F_{contact} = \begin{cases} 1 & \forall (p^i_{hand} - p^i_{arm}) \cdot n^i_{arm} \leq 0 \\ 0 & \exists (p^i_{hand} - p^i_{arm}) \cdot n^i_{arm} > 0 \end{cases} \quad (1)$$

Physical contact attributes were calculated when hand-arm contact was detected. Indentation depth is measured as Equation (2). In particular, $N_C$ is the number of hand vertex points contacted with the forearm. For each contacted hand point $p^i_{hand}$, its indentation depth $d^i$ is approximated as half the distance between $p^i_{hand}$ and its nearest arm point $p^i_{arm}$. The half scale was used because the line between two points might not be perpendicular to the arm surface. The overall indentation $d$ deployed by the hand to the forearm is defined as the average indentation depth of all $N_C$ contacted hand points:

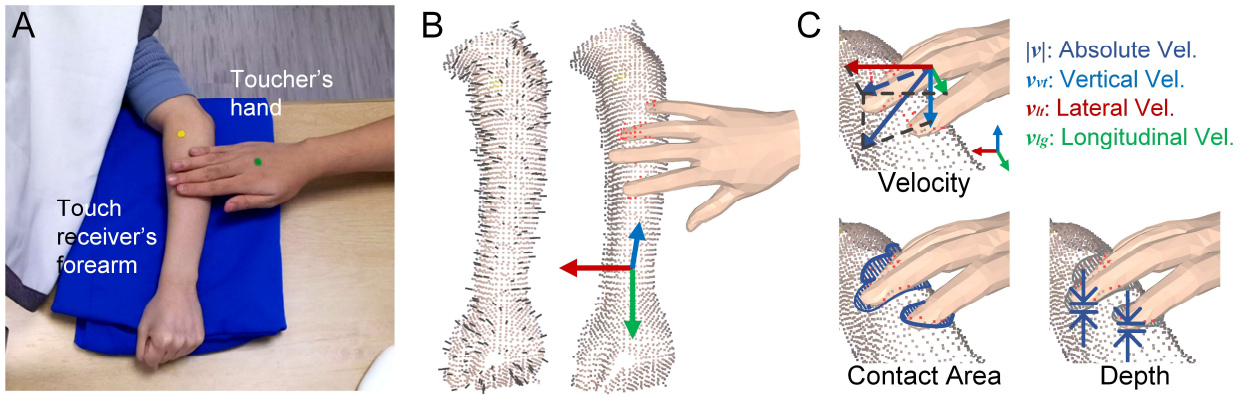

**Figure 2.** Definition of contact attributes. (**A**) Color image from video recorded by depth camera. Two color markers were placed on the toucher's hand and the receiver's forearm respectively to support motion tracking. (**B**) 3D forearm point cloud and hand mesh. Short black line segments represent the norm vector of arm points; red points on the forearm represent the region contacted by the hand. In the arm coordinate, the vertical axis (blue) is designated along the vertical direction pointing right upward, the longitudinal axis (green) is parallel with the arm direction from elbow to wrist, and the lateral direction is perpendicular to the two axes pointing to the internal side of the forearm. (**C**) Six time-series attributes include absolute velocity, which is the absolute value of spatial contact velocity; three orthogonal velocity components corresponding to the three axes of the arm coordinate; contact area, which is the overall area on the forearm being contact; and the indentation depth as the average depth applied on the forearm by the hand.

$$Depth = \frac{\sum_{i=1}^{N_C} \|\boldsymbol{p}_{hand}^i - \boldsymbol{p}_{arm}^i\|_2}{2N_C} . \quad (2)$$

Contact area is measured as the summed area of all contacted arm points. As shown in Equation (3), the unit area $S^i$ for one arm point is calculated as a sphere whose radius is the average neighbor distance, and $\pi$ is round to 3. Within the arm point cloud of $N_{all}$ points, the average neighbor distance $l_{nbr}^i$ is calculated as the average distance of all points to their nearest neighbor points:

$$Area = 3N_C \left(\frac{\sum_{i=1}^{N_{all}} l_{nbr}^i}{N_{all}}\right)^2 . \quad (3)$$

In addition to cutaneous contact attributes, the velocity of hand movement was quantified when contact was detected. The absolute contact velocity $V_{abs}$ is measured as the modulus of the spatial hand velocity $\boldsymbol{v}_{Hand}$:

$$V_{abs} = \left|\frac{\boldsymbol{p}_{Hand}^t - \boldsymbol{p}_{Hand}^{t-1}}{\triangle t}\right| . \quad (4)$$

In Equation (4), hand position $\boldsymbol{p}_{Hand}$ is represented by the position of the middle metacarpophalangeal joint. By defining another coordinate on the receiver's forearm (Figure 2C), spatial hand velocity $\boldsymbol{v}_{Hand}$ is further decomposed in the arm coordinate as three velocity components $V_{vt}, V_{lg}, V_{lt}$ parallel with its axis of the arm coordinate (Figure 2C). The vertical axis $\boldsymbol{i}_{vt}$ of the arm coordinate is aligned with the vertical direction pointing upright. It could be obtained as the normal vector of a point on a horizontal surface, like a table, or the normal vector of a point on the top of the receiver's forearm. Vertical velocity $V_{vt}$ is the hand velocity component in this direction:

$$V_{vt} = \boldsymbol{v}_{Hand} \cdot \boldsymbol{i}_{vt} . \quad (5)$$

The longitudinal axis $\boldsymbol{i}_{lg}$ is aligned with the direction of the arm bone, pointing from elbow to wrist. To derive this axis, the camera was orientated to display the forearm vertically in the 2D image. Then, the direction of the arm bone in the 2D image was set to be parallel with the y axis of the image coordinate. By projecting the y axis $\boldsymbol{y}$ of the camera coordinate onto the perpendicular plane of the vertical axis $\boldsymbol{n}_{vt}$, the longitudinal axis follows the direction of the projected vector:

$$\boldsymbol{i}_{lg} = \frac{\boldsymbol{y} - (\boldsymbol{y} \cdot \boldsymbol{i}_{vt})\boldsymbol{i}_{vt}}{\|\boldsymbol{y} - (\boldsymbol{y} \cdot \boldsymbol{i}_{vt})\boldsymbol{i}_{vt}\|_2} . \quad (6)$$

$$V_{lg} = \boldsymbol{v}_{Hand} \cdot \boldsymbol{i}_{lg} . \quad (7)$$

Lastly, the lateral axis $\boldsymbol{i}_{lt}$ is perpendicular to the plane of longitudinal and vertical axis, following the right-hand rule:

$$\boldsymbol{i}_{lt} = \boldsymbol{i}_{lg} \times \boldsymbol{i}_{vt} . \quad (8)$$

$$V_{lt} = \boldsymbol{v}_{Hand} \cdot \boldsymbol{i}_{lt} . \quad (9)$$

Compared with the overall hand velocity, these velocity components can quantify the directional nature of the hand movements.

Moreover, contact duration is measured as a scalar value for each hand-arm touch interaction, which is the sum of time over which contact was detected. Given the recording frequency $f$ of the camera is 30 Hz and $N_f$ is the number of frames per interaction, the contact duration is measured as:

$$Duration = \frac{\sum_{i=1}^{N_f} F_{contact}}{f} \quad . \quad (10)$$

## 3 Experiment 1: Human-to-Human Affective Touch Communication

The first experiment was designed with the task of human-to-human emotion communication. Touchers was instructed to deliver cued emotional messages, e.g., happiness, sympathy, anger, to the touch receiver at the receiver's forearm using preferred gestures, e.g., tapping, holding, stroking. Recorded contact attributes were then used to differentiate delivered messages, utilized gestures, and individual touchers. Contact analysis was conducted on the platform with the Intel Core i9-9900 CPU, 3.1 GHz, 64 GB RAM, and a NVIDIA GeForce RTX 2080 SUPER GPU. The same platform was used for the second experiment.

### 3.1 Cued Emotional Messages and Gesture Stimuli

Seven emotions of anger, attention, calm, fear, gratitude, happiness, and sympathy were selected as cued messages for touchers to express (Table 1). Those messages were adopted from prior studies and have been observed to be recognizable through touch alone (Hertenstein et al., 2006, 2009; Thompson and Hampton, 2011; Hauser et al., 2019a; McIntyre et al., 2021). Among them, gratitude and sympathy are prosocial expressions that are more effectively communicated by touch compared with those self-focused. Anger, happiness, and fear are universal expressions that are commonly communicated by facial, vocal, and touch expressions. Attention and calm are also preferred messages in touch interactions and can be correctly interpreted significantly better than chance. For each of the cued messages, three commonly used gestures were adopted from prior studies (Hertenstein et al., 2006; Thompson and Hampton, 2011; Hauser et al., 2019a; McIntyre et al., 2021) (Table 1). Holding and squeezing were combined into one since they share a similar hand gesture and hand motion. Similarly, hitting was combined with the tapping gesture, but only for the message of anger.

Table 1. Available gestures for each cued emotional message in touch communication task

| | Cued Emotional Messages | | | | | | |
|---|---|---|---|---|---|---|---|
| | Anger (Ag) | Attention (At) | Calm (C) | Fear (F) | Gratitude (G) | Happiness (H) | Sympathy (S) |
| **Gestures** | Hit/Tap | Tap | Hold/Squeeze | Squeeze/Hold | Hold/Squeeze | Shake | Stroke |
| | Squeeze/Hold | Shake | Stroke | Shake | Shake | Tap | Tap |
| | Shake | Squeeze/Hold | Tap | Tap | Tap | Stroke | Squeeze/Hold |

### 3.2 Participants

The human-subjects experiments were approved by the Institutional Review Board at the University of Virginia. Ten participants were recruited as touchers, including five males and five females (mean age = 23.8, SD = 5.0). Another five participants were recruited as touch receivers with three males and two females (mean age = 24.0, SD = 4.4). Five experimental groups were randomly assembled, where each group consisted of one male toucher, one female toucher, and one receiver. Each group performed two experimental sessions with one session conducted by the male toucher and another one conducted by the female toucher. Written informed consent was obtained from all participants.

### 3.3 Experimental Setup

To avoid visual distractions during the experiment, touchers and receivers sat at opposing sides of an opaque curtain. They were instructed to not speak to each other. As shown in Figure 2A, a cushion was set on the table at the toucher's side upon which the receiver rested her or his left forearm. Cued emotional messages and corresponding gestures were displayed to the toucher on the computer screen. The toucher could select the gesture and proceed to the next message using the computer's mouse. Cued messages and the toucher's selection of gestures were also recorded. As illustrated by a snapshot of the experiment recoding by depth camera (Figure 2A), the camera was set in front of the cushion and orientated towards it.

### 3.4 Experimental Procedures

In each session, seven cued emotional messages were communicated with each repeated six times. The 42 message instructions were provided in random order. In each trial, one message was displayed on the screen with three gestures listed below. Touchers had five seconds to choose a gesture and report it on the computer display. For each cued message, the three provided gestures were identical but their order was randomized trial by trial. After that, the toucher delivered the message, by touching the receiver's forearm from elbow to wrist, using the right hand. Within each trial, only the chosen gesture was used. The use of other gestures or a combination of gestures was not allowed. For the same cued message across trials, touchers were free to use the same gesture or change to another gesture. A gesture could be deployed in any pattern of contact deemed appropriate by the toucher. No constraints or instructions were given for delivering the gesture, such as its duration, hand region employed, intensity, or repetition. At the end of a trial, by clicking the 'Next' button on the bottom of the computer display, the toucher initiated the next trial with a new message word and corresponding three gestures.

### 3.5 Data Analysis

Overall, 420 trials were performed in ten experimental sessions. Twelve trials were excluded from analysis as contact interactions were not properly recorded. Statistical and machine learning analyses were performed to examine the measured contact attributes.

To identify the contact pattern between touch gestures, paired-sample Mann–Whitney U tests were applied across gestures per contact attribute. For time-series attributes, the mean value was used. Since longitudinal velocity, lateral velocity, and vertical velocity are signed variables, the mean was derived from the absolute value of those variables. Contact duration as a scalar variable was directly compared across gestures. To evaluate which of the contact attributes could best identify or describe a certain type of touch gesture, the importance of each attribute in predicting that gesture was identified using a random forest classifier. The mean values of time-series attributes together with the scalar attribute served as inputs. For example, in predicting the stroking gesture, all trials were labeled in a binary fashion as delivering or not delivering this gesture, instead of being labeled as the four gesture types. Seventy-five percent of trials were randomly assigned as the training set and those remaining were assigned as the test set. The permutation method was used to derive the importance of attributes. The value was obtained as the average of 100 repetitions of classification, with 10 permutations per classification.

Further classification analyses were performed regarding the discrimination of touch gestures, emotional messages, and individual touchers, respectively, using the random forest algorithm. Contact attributes were fed into classifiers in three different formats, including the mean value of each time-series attribute, multiple relevant features extracted from each time-series attribute, and the original time-series attributes. In particular, multiple features were extracted to quantify the amplitude, frequency, and dynamic characteristics of the time-series signal (Christ et al., 2018). For example, time-domain features included mean, maximum, quartiles, standard deviation, trend, skewness, entropy, energy, etc. Frequency domain features included autocorrelations and partial autocorrelations with different lags, coefficients of wavelet and Fourier transformations, mean, variance, skew of Fourier transform spectrum, etc. From all extracted features, relevant ones were selected for classification by significance tests in predicting the classification target and the Benjamini Hochberg multiple test (Christ et al., 2018). When time-series data were used, all attributes were concatenated into one variable as input (Löning et al., 2019). To identify attributes that could better encode social affective touch, the importance of individual attributes was ranked for each classification task. More specifically, based on the mean-value classification, the permutation method was repeated multiple times to derive the average importance values.

### 3.6 Results

#### 3.6.1 Physical Contact Attributes in Human-to-Human Touch

Human-to-human physical contact interactions between social messages, gestures, and individual touchers were quantified by their contact attributes. As shown in Figure 3, exemplar data for the four touch gestures (shake, tap, hold and stroke) exhibit distinct patterns across the contact attributes, consistent with expected hand movements per gesture. In particular, the stroking gesture was characterized by regular patterns in longitudinal velocity, which implies slow and repetitive movements along the direction of the forearm. For the shaking gesture, velocity attributes depicted large changes in frequency and relatively lower amplitude. Meanwhile, velocities in all three directions changed simultaneously, indicating a spatial direction in the movement of the toucher's hand. The tapping gesture was quantified as discontinuous, large-amplitude spikes of short contact duration. Compared with other touch gestures, holding gesture exhibited relatively stable contact with minimal changes. With further inspection into each gesture, contact patterns with subtle differences could also be captured across emotional messages. Such as in the shaking gesture, happiness was delivered with higher velocities compared with the expression of fear. Within the tapping gesture, shorter but more intensive contact was recorded when expressing anger compared with attention.

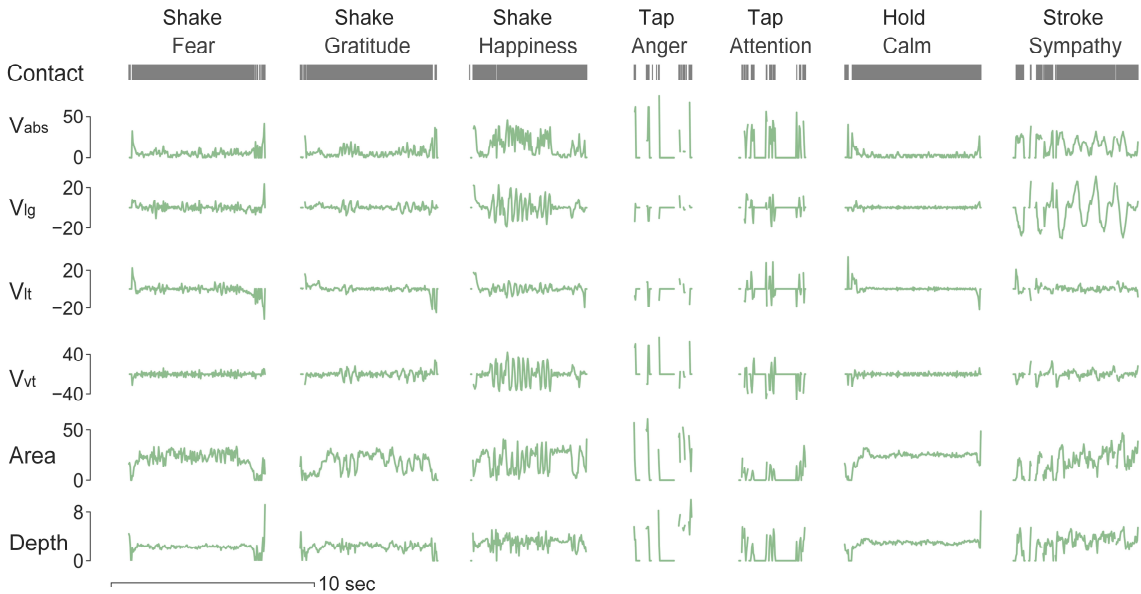

**Figure 3.** Time-series recordings of each contact attribute across touch gestures and delivered messages. Distinct contact patterns were captured by the spatiotemporal changes of those attributes. The *Contact* variable represents the status of the being contacted or not. $V_{abs}$ denotes the absolute contact velocity (cm/s), $V_{lg}$ denotes the longitudinal velocity (cm/s), $V_{lt}$ denotes the lateral velocity (cm/s), $V_{vt}$ denotes the vertical velocity (cm/s), *Area* denotes the contact area (cm$^2$), and *Depth* denotes the indentation depth (mm).

As shown in Figure 4A, the four touch gestures were statistically differentiable according to several of their contact attributes. For instance, absolute contact velocity can differentiate all gesture pairs except for that of stroking and shaking. With the contact attribute of longitudinal velocity, stroking was differentiable from shaking as it afforded higher longitudinal velocity. This also aligns with hand movements during stroking that are typically along the direction of the forearm. Both shaking and tapping gestures exhibited significantly higher longitudinal velocities than the holding gesture. With the lateral velocity, significant differences were derived among all four gestures, where tapping and shaking gestures afforded higher amplitudes than stroking and holding. As for the vertical velocity, the tapping gesture was associated with significantly higher velocities than others, which aligns with its up-down movements. Across all velocity attributes, the holding gesture was significantly distinct from other ones.

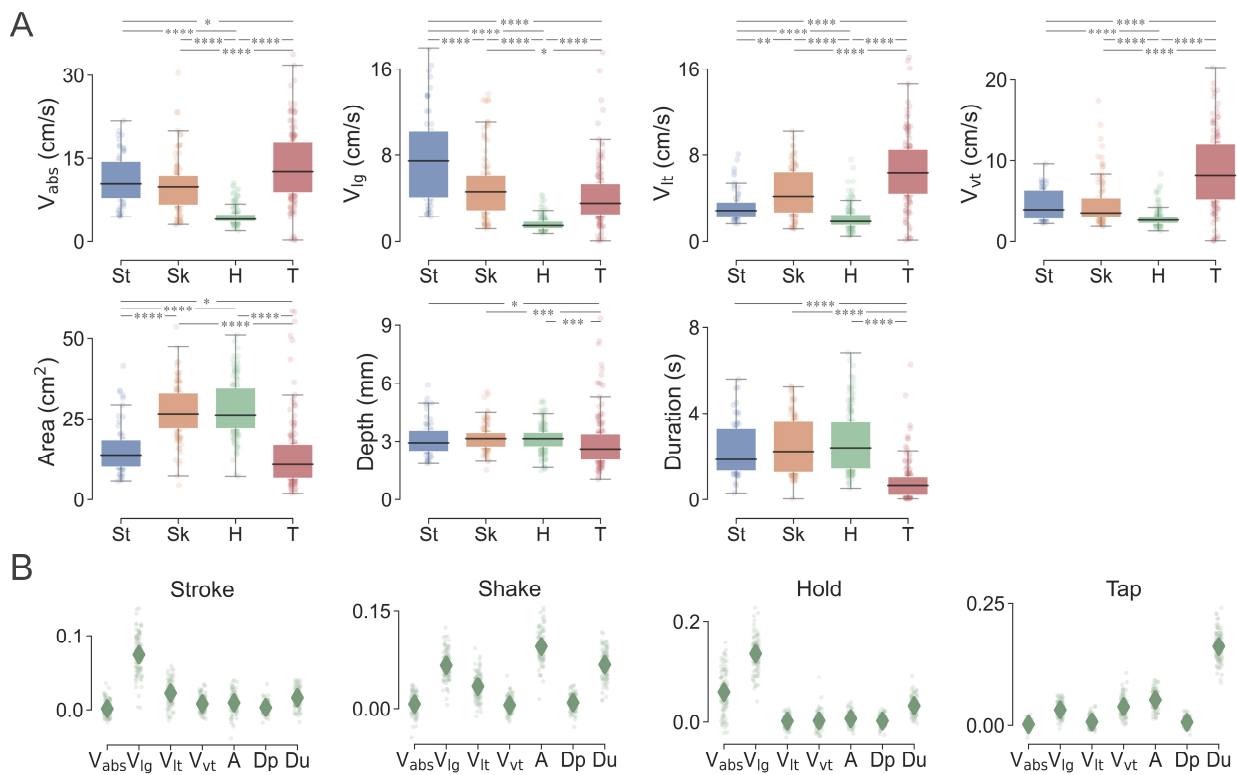

**Figure 4.** (**A**) Comparison of contact attributes across the four touch gestures. *$p < 0.05$, **$p < 0.01$, ***$p < 0.001$, ****$p < 0.0001$ were derived by paired-sample Mann–Whitney U tests. (**B**) Importance of certain contact attributes in identifying each touch gesture using random forest classification. Diamonds denote means; points denote importance values of 100 repetitions of classification.

For the contact area attribute, shaking and holding gestures exhibited significantly higher values than the stroking gesture, and then tapping. Indeed, participants generally used the whole hand to deliver holding and shaking, while only the finger digits for stroking and the fingertips for tapping. Moreover, with indentation depth and contact duration, tapping was distinct amongst the gestures with significantly lower depth and shorter duration. Note the hand motion with the tapping gesture could be faster than the recording frequency of the camera, where one trial of contact might not be entirely captured and thus lead to a lower estimation of indentation depth.

In Figure 4B, the contact attributes that were salient in identifying or describing a specific touch gesture were further analyzed according to their importance in predicting that gesture. From the importance ranking, longitudinal velocity appears to be the most useful attribute in describing the stroking gesture. The shaking gesture did not have a single salient attribute, perhaps because it was delivered from multiple directions and varied velocities. The attributes of contact area, contact duration, and longitudinal velocity were relatively more important. The holding gesture could be identified by longitudinal and absolute velocities with both lower amplitudes. For the tapping gesture, contact duration could be important in identifying it, which should be shorter than other gestures.

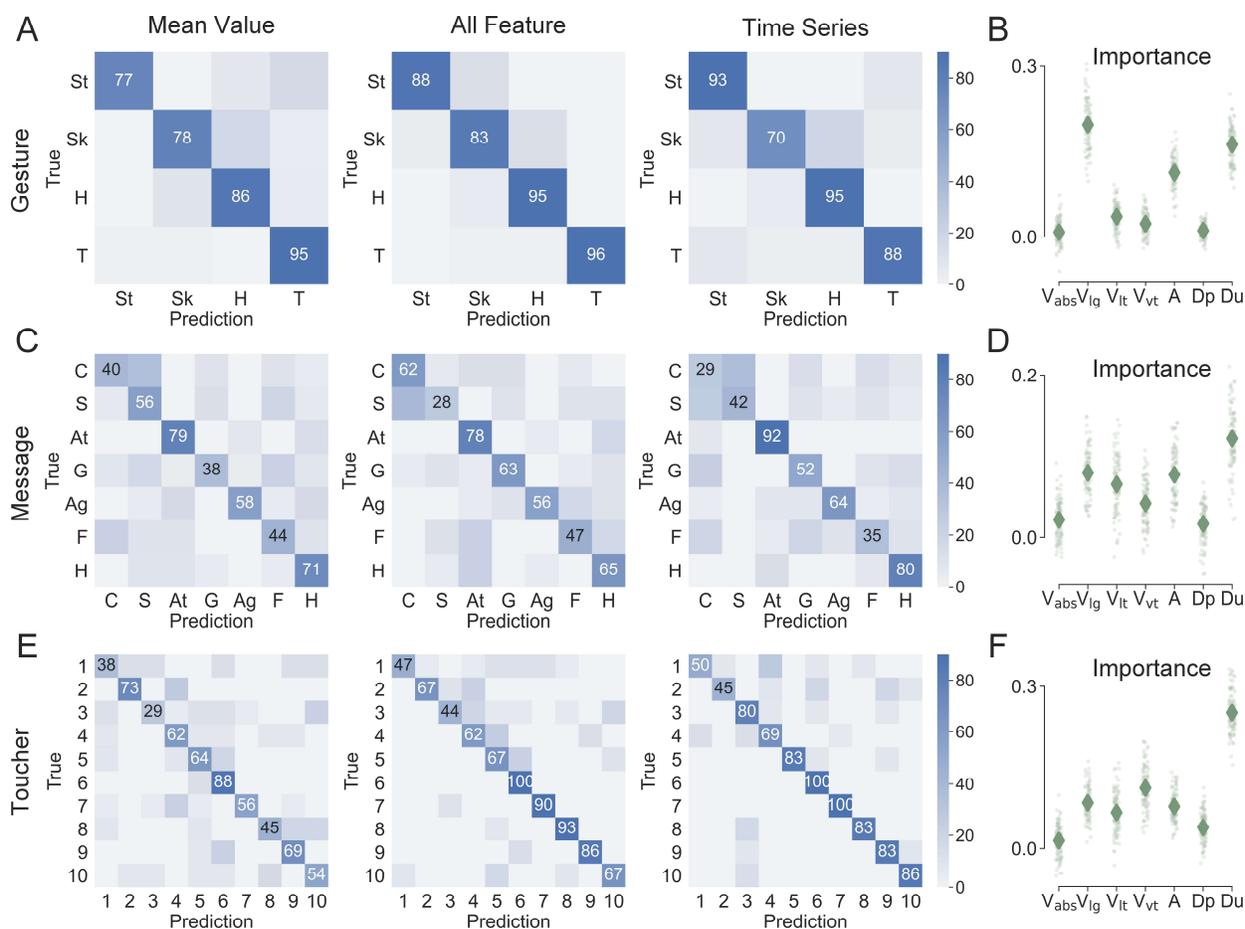

**Figure 5.** Classification of touch gestures, delivered messages, and toucher individuals using the mean value, all relevant features, and time-series data of contact attributes, respectively. The accuracy in prediction of (**A**) touch gestures, (**C**) delivered messages, (**E**) toucher individual are shown, as well as the importance of particular contact attributes in classifying (**B**) touch gestures, (**D**) delivered messages, (**F**) toucher individual. Numbers and colors in confusion matrices represent the prediction percentage. In the importance plots, the diamonds denote means; points denote importance values from 100 repetitions of classification.

### 3.6.2 Classification amidst Gestures, Messages, and Individuals

In Figure 5, the contact attributes are shown to robustly classify touch gestures, delivered messages, and individual touchers at accuracies better than chance, which is 25%, 14.3%, and 10% respectively. For gesture prediction, the accuracy was 87% when the mean values of contact attributes were used as predictors (Figure 5A). The prediction accuracy slightly increased to 92% when all relevant features were used as more information was included, and was around 86% when predicted by the time-series data. In classifying delivered emotional messages, the accuracy was 54%, 57%, and 55%, for the three respective feature classes (Figure 5C). Moreover, in classifying the individual touchers, the accuracies were 56%, 72%, and 77%, respectively. For the importance ranking of the contact attributes, those of longitudinal velocity, contact duration, and contact area were typically more important.

## 4 Experiment 2: Technical Validation on the Visual Tracking Method

The second experiment was designed to validate the effectiveness of the 3D visual tracking system in measuring controlled human movements against those from independent devices, including an electromagnetic motion tracker, sensorized pressure mat, and laser displacement sensor. These techniques are used commonly in haptics studies (Silvera-Tawil et al., 2014; Jung et al., 2015; Hauser et al., 2019a; Xu et al., 2020, 2021a; Lo et al., 2021). In this experiment, the observed contact attributes were compared within controlled touch conditions, e.g., stroking in different directions at preset velocities, pressing with different parts of the hand varying in contact area, and tapping at different depth magnitudes.

### 4.1 Contact Velocity Validation Using Electromagnetic Tracker

#### 4.1.1 Experimental Setup

Measurements of the directional components of contact velocity, including absolute velocity, longitudinal velocity, lateral velocity, and vertical velocity were validated against those of an electromagnetic (EM) motion tracker (3D Guidance, Northern Digital, Canada. 6 DOF, 20-255 Hz, 1.4 mm RMS position accuracy, 78 cm range; 0.5° RMS orientation accuracy, ±180° azimuth & roll, ±90° elevation range). Both tracking systems were operated simultaneously to capture controlled movements of the human hand touching the forearm. The transmitter of the 3D Guidance EM tracker was oriented to be aligned with the arm coordinate (Figure 6A). The sensor of the EM tracker was attached to the toucher's back of the hand near the middle metacarpophalangeal joint.

#### 4.1.2 Experimental Procedures

Given velocity components were defined in different directions, five test gestures were designed in total. The first two test gestures were stroking contact along the forearm in longitudinal and lateral directions, respectively. The third test gesture involved tapping vertically to the surface of the forearm. The fourth gesture was holding without movement. The fifth gesture was shaking, which was delivered in an irregular and arbitrary way with different directions and velocities included. For the first three test gestures, each one was performed in three levels of velocities, from low to medium to high. Each velocity level was repeated for three trials. For example, the longitudinal stroking gesture was performed as three trials of stroking in the longitudinal direction with lower velocity, followed by three trials of stroking with medium velocity, and concluded by three trials of stroking with higher velocity. The direction of hand movement and level of velocity were behaviorally controlled by the trained toucher, who performed all three validation experiments. Shaking and holding gestures were performed only once but lasted for a longer time to collect enough amount of data for validation analysis.

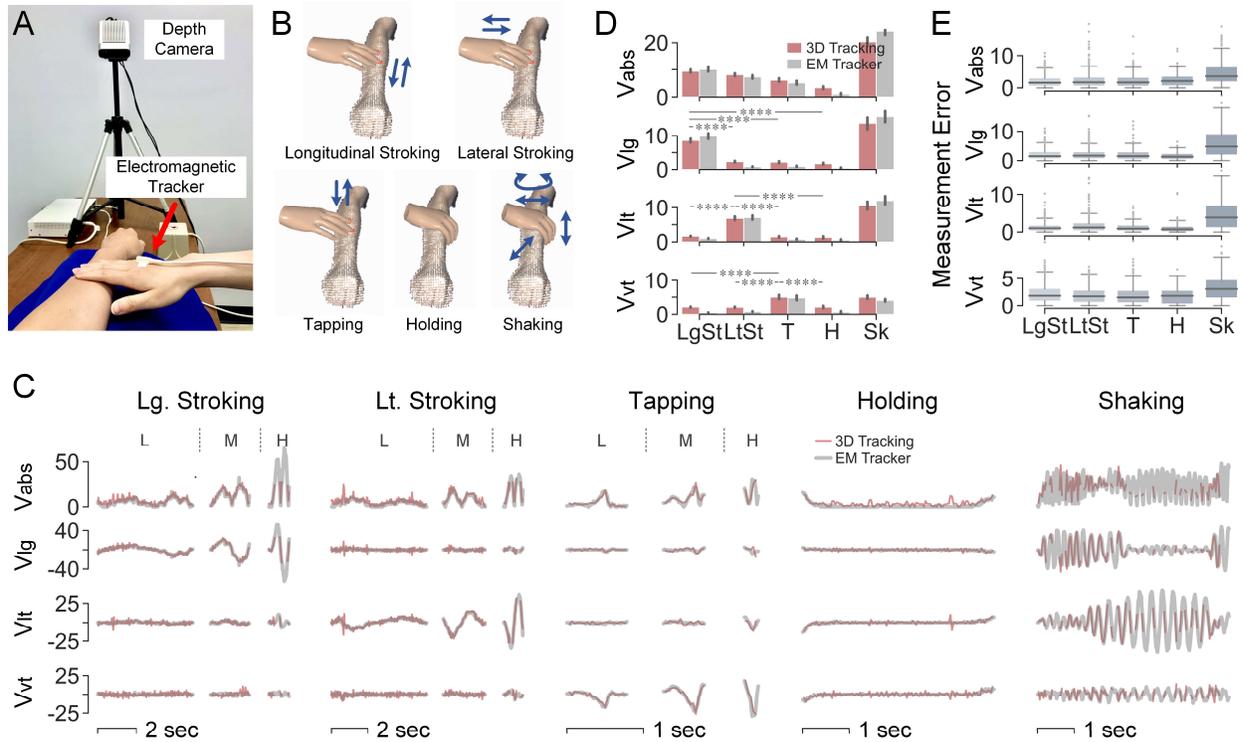

**Figure 6.** Validation of contact velocity measurements using EM tracker. (**A**) Experimental setup. (**B**) Five test gestures. (**C**) Velocity (cm/s) over time by the two tracking systems. For the first three test gestures, one trial is shown per force level, i.e., low, medium, and high force. (**D**) Mean values of velocities (cm/s) per test gesture. ****$p < 0.0001$ were derived by paired-sample Mann–Whitney U tests. (**E**) Errors (cm/s) of measured velocities between the two systems for each test gesture.

**Table 2.** Experiment procedure for validating contact velocity

| | Test Gesture | Moving Direction | Velocity Levels | Repeated Trials per Level | Trials in Total |
|---|---|---|---|---|---|
| 1 | Stroking | Longitudinal | Low, Medium, High | 3 | 9 |
| 2 | Stroking | Lateral | Low, Medium, High | 3 | 9 |
| 3 | Tapping | Vertical | Low, Medium, High | 3 | 9 |
| 4 | Holding | None | None | 1 | 1 (long duration) |
| 5 | Shaking | Irregular | Irregular | 1 | 1 (long duration) |

### 4.1.3 Data Analysis

Similar to the 3D visual tracking system, the four velocity attributes captured by the EM tracker were derived from the original time-series position data. For either tracking system, the absolute mean value of each velocity attribute was calculated per test gesture. Mann–Whitney U tests were conducted across the test gestures based on mean velocity collected by the visual tracking system. Measurement errors between the two tracking systems were derived per attribute and test gesture. Since the sampling rates of the two systems differ, i.e., 30 Hz for the Azure Kinect camera and 60 Hz for the EM tracker, data collected from the EM tracker was resampled to be synchronized. More specifically, the EM tracking data was first interpolated and sampled according to the timestamps of the 3D visual tracking data. Then, the error was calculated for each time point between the velocities from the two systems.

### 4.1.4 Results

In Figure 6, velocities measured by the 3D visual tracking system were accurate when compared with the EM tracker. The time-series data from the two systems well overlaped amidst touch gestures (Figure 6C) and the average velocities of the gestures were comparable between the two systems (Figure 6D). Shaking delivered high velocities in all three directions, while velocity in a certain direction was significantly higher for hand movements along that direction. All four velocity attributes were significantly lower when the holding gesture was performed. As shown in Figure 6E, the measurement error was 1-2 cm/s for the first four gestures and relatively higher at around 5 cm/s for the shaking gesture.

### 4.2 Contact Area Validation Using Sensorized Pressure Mat

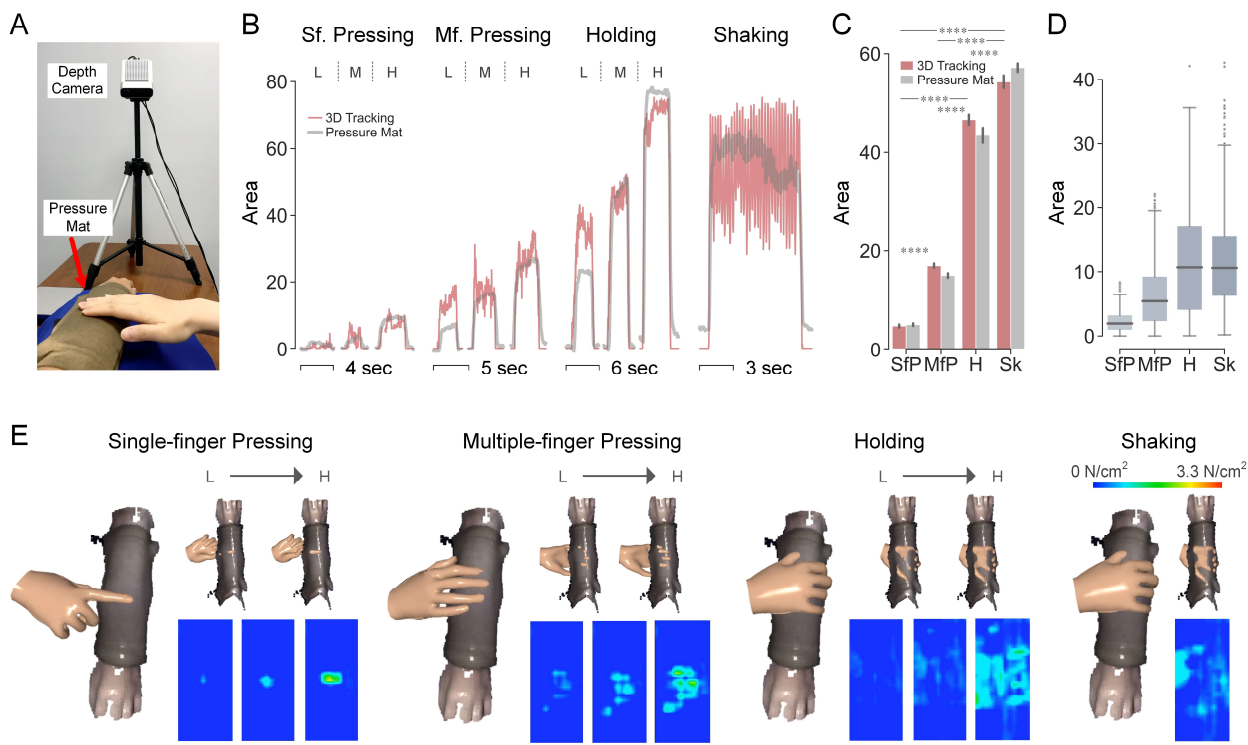

**Figure 7.** Validation of contact area measurements using sensorized pressure mat. (**A**) Experimental setup. (**B**) Contact area (cm$^2$) over time by the two systems. For the first three test gestures are shown one trial per force level, i.e., low, medium, and high force. (**C**) Mean values of contact area (cm$^2$) per test gesture. ****$p < 0.0001$ were derived by paired-sample Mann–Whitney U tests. (**D**) Differences of measured contact area (cm$^2$) between the two systems per test gesture. (**E**) Visualization of hand-arm contact in top view (left) and bottom view (top right) with heatmaps of contact pressure tracked by sensorized pressure mat across force levels (bottom right).

### 4.2.1 Experimental Setup

Contact area was measured simultaneously with the 3D visual tracking system and a sensorized pressure mat (Conformable TactArray SN8880, Pressure Profile Systems, USA, 7x14 cm, 12x27 sensing elements, 0.002 psi pressure resolution, 3.05 psi pressure range, 29.3 Hz). Note that contact was evaluated between the toucher's hand and the surface of the pressure mat which was overlaid on top of the bare forearm, for which it had been custom-designed (Figure 7A). Based on pilot tests with the pressure mat, its measurement of contact area could be inaccurate due to the creases caused by pressing when the mat was put on the forearm. To attenuate this effect, a piece of single-face corrugated cardboard was placed between the forearm and the mat to generate a smooth and stiffer curved surface following the shape of the forearm.

### 4.2.2 Experimental Procedures

Four test gestures were employed. The first test gesture was single-finger pressing with the index finger. The second gesture was multiple-finger pressing with all fingers except for the thumb. The third gesture was holding and the fourth gesture was shaking. For the first three test gestures, three levels of force were applied from low to medium to high, to generate different levels of contact area within a gesture. Each force level was repeated for three trials. Per trial, the toucher's hand moved downward into the receiver's forearm and maintained pressure/hold at that force level for more than three seconds. For example, the single-finger pressing gesture was conducted for three trials of pressure using the index finger at a low force level, followed by three trials of pressure at a medium force level, and three trials of pressing with a higher force level. The shaking gesture was conducted for one trial with a long duration. Any patterns of shaking could be applied in an irregular and arbitrary manner including different directions, velocities, etc.

Table 3. Experiment procedure for validating contact area

|   | Test Gesture | Force Levels | Repeated Trials per Level | Trials in Total |
|---|---|---|---|---|
| 1 | Single-finger pressing | Low, Medium, High | 3 | 9 |
| 2 | Multiple-finger pressing | Low, Medium, High | 3 | 9 |
| 3 | Holding | Low, Medium, High | 3 | 9 |
| 4 | Shaking | Irregular | 1 | 1 (long duration) |

### 4.2.3 Data Analysis

The average contact area per gesture was calculated for both measurement systems. Significance tests were performed across gestures based on average areas from the visual tracking system. The measurement differences between the two systems were derived from time-series recordings per gesture. To overcome the time discrepancy of sampling, data collected by the sensorized pressure mat was resampled to be synchronized with the visual tracking system.

### 4.2.4 Results

In Figure 7B, the time-series contact areas captured by the 3D visual tracking system and the sensorized pressure mat well overlapped with each other across test gestures and force levels. While single-finger pressing (SfP) afforded the smallest contact area, larger multiple-finger pressing (MfP) was significantly smaller than holding (H) and shaking (Sk) (Figure 7C). As shown in Figure 7D, the measurement differences between the two systems were around 2 and 6 $cm^2$ for SfP and MfP, while increased to 11 $cm^2$ for holding and shaking.

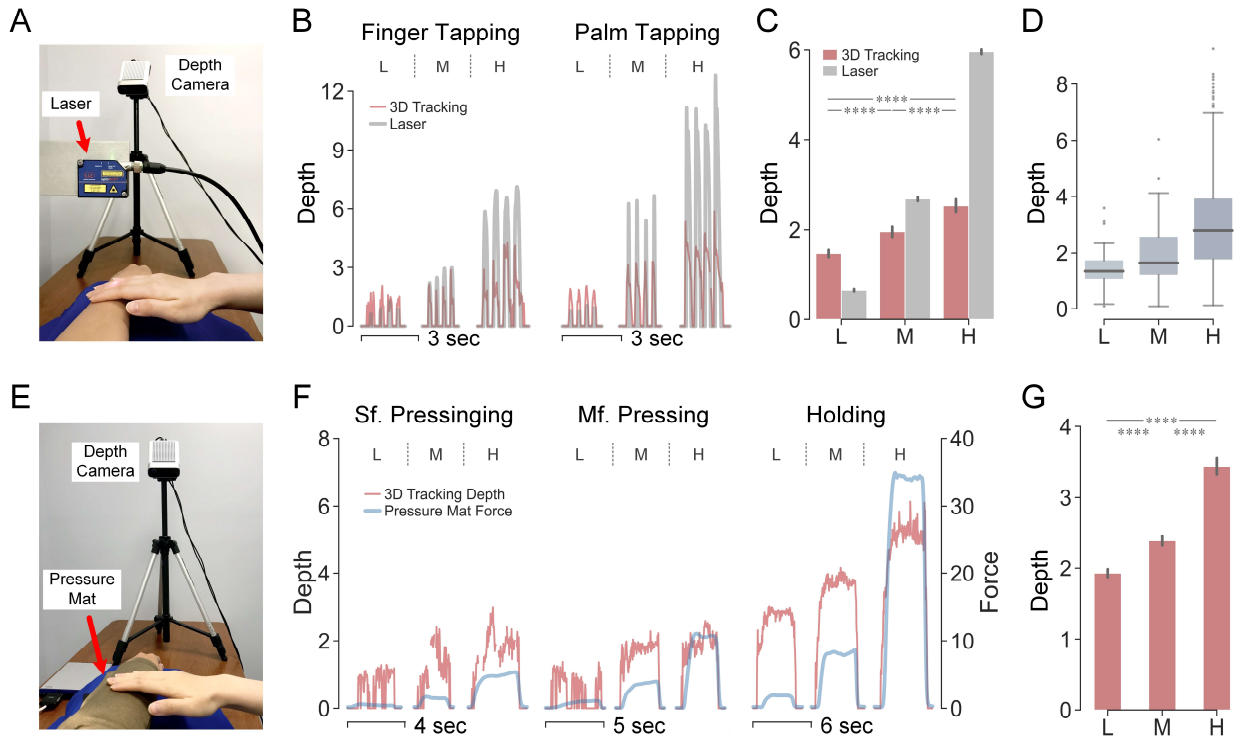

**Figure 8.** Validation of indentation depth measurements using laser displacement sensor and sensorized pressure mat. (**A**) Experimental setup with laser displacement sensor. (**B**) Indentation depth (mm) over time by the either system. For the two test gestures shown is one trial per force level, i.e., low, medium, and high force. (**C**) Mean values of indentation depth per test gesture. ****$p < 0.0001$ were derived by paired-sample Mann–Whitney U tests across force levels. (**D**) Errors of measured indentation depth between systems per force level. (**E**) Experimental setup with sensorized pressure mat. (**F**) Indentation depth (mm) collected by the 3D visual tracking system overlaps with overall force (N) collected by the sensorized pressure mat. Per test gesture, one trial per force level is shown i.e., low, medium, and high force. (**G**) Mean value of indentation depth per force level recorded by the 3D visual tracking system. ****$p < 0.0001$ were derived by paired-sample Mann–Whitney U tests across force levels.

### 4.3 Indentation Depth Validation Using Laser Sensor

#### 4.3.1 Experimental Setup

Indentation depth was first validated using a laser displacement sensor (optoNCDT ILD 1402-100, Micro-Epsilon, Germany, 100 mm range, 10 µm resolution, 1.5 kHz). The sensor was mounted on a customized stand with the beam pointing downward. Given its capability of measuring the displacement of one point in only the vertical direction (Figure 8A), a limited set of tapping gestures was evaluated in this setting. Other gestures were then tested with a separate validation procedure using the sensorized pressure mat (Figure 8E).

#### 4.3.2 Experimental Procedures

Two test gestures were examined with the laser sensor. The first gesture was multiple-finger tapping, where the movement of the tip of the middle finger was tracked. The second gesture was tapping with the palm, measured at one point on the back of the hand. Holding, shaking, and stroking gestures were not examined here since these gestures are typically not conducted in the vertical direction. Within each gesture, three force levels were employed, i.e., low, medium, high, and each repeated in three trials. The toucher quickly tapped for four times within one trial. For example, the palm tapping gesture was conducted for three trials of four taps with the palm at a low force level, followed by three trials of four taps at a medium force level, and three trials of four taps at a high force level. The raw data collected by laser sensor contained displacements of both indentations into the skin and movements in the air. Therefore, the toucher conducted a 'zero contact' touch to the forearm at a minimally perceptible force prior to each test gesture.

Within the setting of sensorized pressure mat, the three test gestures performed were single-finger pressing, multiple-finger pressing, and holding. Each gesture was performed in three force levels, where each level was repeated for three trials.

Table 4. Experiment procedure for validating indentation depth

| | Test Gesture | Force Levels | Repeated Trials per Level | Trials in Total |
|---|---|---|---|---|
| | **Validation with Laser Sensor** | | | |
| 1 | Multiple-finger tapping | Low, Medium, High | 3 (4 taps per trial) | 9 |
| 2 | Palm tapping | Low, Medium, High | 3 (4 taps per trial) | 9 |
| | **Validation with Pressure Mat** | | | |
| | Test Gesture | Force Levels | Repeated Trials per Level | Trials in Total |
| 1 | Single-finger pressing | Low, Medium, High | 3 | 9 |
| 2 | Multiple-finger pressing | Low, Medium, High | 3 | 9 |
| 3 | Holding | Low, Medium, High | 3 | 9 |

### 4.3.3 Data Analysis

For the validation with laser sensor, average indentation depth at each force level was obtained by aggregating the two tapping gestures. Significance tests were conducted across force levels based on the average depth collected by the visual tracking system. Measurement errors between the two systems were derived from time-series recordings at each force level. The data from the laser sensor was resampled according to the 3D visual tracking system's results. For quick tapping gestures, slight temporal discrepancies between the two recordings could derive large differences. Therefore, the dynamic time warping method was used to match tracked movements. The measurement errors were obtained by comparing each pair of matched points from the two recordings.

Though no depth data could be captured by the pressure mat, the overall contact force was measured for correlation with indentation depth measured by the visual tracking system. By aggregating all test gestures, the average depth derived per force level was then calculated and compared.

### 4.3.4 Results

In Figure 8, the patterns of indentation depth measured by the two systems were very similar especially for the temporal changes (Figure 8B). Though differences could be observed between their overall amplitudes, their increasing trends were maintained across force levels (Figure 8C). Therefore, the 3D visual tracking system affords the sensitivity to track slight changes in indentation depth, while the amplitude of changes is proportionally mitigated. Moreover, contact with different force levels could be easily differentiated by indentation depth amongst a variety of touch gestures. (Figure 8C, 8G).

## 5    Discussion

To better understand human-to-human touch interactions underlying social emotional communication, an interference-free 3D visual tracking system was developed to precisely measure skin-to-skin physical contact by time-series contact attributes. The system was validated to capture and readily distinguish naturalistic human touches across delivered emotional messages, touch gestures, and individual touchers according to contact attributes. Compared with standard tracking techniques, similar accuracy of spatiotemporal measurements was achieved by this system, while multivariate attributes can be obtained simultaneously within one concise setup.

## 5.1 Deciphering Affective Touch Communication by Contact Attributes

As human affective touch is prone to be impacted by social and individual factors, such contact differences could be readily captured by this system via contact attributes. First of all, touch gestures can be differentiated with high accuracy as their contact attributes were significantly different from each other (Figure 4A). Measurements of this system also align with prior reports of gesture quantification with similar amplitudes. Such as the velocity for stroking in social touch is around 10 cm/s (Lo et al., 2021), and the average contact area of holding gesture is around 30 $cm^2$ (Hauser et al., 2019a). In addition, the characterized contact pattern of each gesture align well with the general sense of how we deliver that gesture. For example, tapping is associated with higher vertical velocities, stroking is delivered with higher longitudinal velocities, and holding is commonly applied with lower velocities and larger contact areas (Figure 4A).

Moreover, delivered emotional messages can be differentiated by contact attributes much better than chance (Figure 5C). The accuracy of 54%, 57%, 55% was achieved when predicted by three different levels of information derived from contact attributes (Figure 5C). Note that human receivers only achieve a comparable recognition correctness around 57% when a similar pool of messages were tested (Hauser et al., 2019a; McIntyre et al., 2021). It indicates that some contact information human receivers rely on in identifying emotional messages can be captured by this tracking system. Meanwhile, certain messages that were difficult to be discriminated by contact attributes might indeed be very similar in their social meanings and touch behaviors. Such as sympathy and calm, which are supposed to be close in the terms of contact quantification.

Furthermore, this tracking system can capture individual differences in affective touch as individual touchers were also easily distinguished. Prior studies highlighted that touch behavior in social communication could be influenced by many factors, such as age (Cascio et al., 2019), gender (Hertenstein et al., 2009; Russo et al., 2020), cultural backgrounds (Hertenstein et al., 2006; Suvilehto et al., 2019), relationship (Thompson and Hampton, 2011), or personalities (McIntyre et al., 2021). While the personal information is easy to obtain via questionnaires, the uniqueness of their contact performance is always challenging to collect. Prior attempts on individual difference typically focused on contact with engineered stimuli like silicone-elastomers (Xu et al., 2021b), grooved surfaces in grating orientation tasks (Peters et al., 2009), or the contact with robots (Cang et al., 2015). In those settings, contact can be well-recorded by built-in or attached sensors, which in contrast is impractical or interferential for human-to-human touch. As individual difference indeed plays a role in social emotion communication, this system could help bridge the gap by inspecting the differences from the aspect of skin contact quantification.

## 5.2 Improved Skin-to-Skin Contact Measurement by 3D Visual Tracking

The measurement accuracy of this system was validated by several standard tracking techniques. As shown in Figures 6-8, time-series recordings of contact attributes aligned well with the data collected from independent devices, i.e., contact velocities from an EM motion tracker, contact area from a sensorized pressure mat, and indentation depth from a laser sensor. Those standard tracking methods typically afford high accuracy or resolution of measurements but are specialized for limited types of contact attributes. Therefore, when different attributes are needed at the same time, a complex combination of multiple devices is usually required. In contrast, the proposed tracking system captures most of those attributes simultaneously with a concise setup without calibration.

Moreover, the proposed 3D visual tracking system is compatible with wider applications as many limitations of standard tracking methods were overcome or avoided. More specifically, compared with the EM tracker, this system is free of electromagnetic interference and provides shape information instead of tracking the position of only few points. Compared with infrared motion trackers like the Leap Motion sensor, it covers a larger range of tracking and captures any 3D shapes in addition to hands and several basic geometric shapes. The motion capture system is superior in tracking movements but is expensive to set up and constrained by pre-attached markers. Sensorized pressure mat and other force sensors always block the direct contact and might not be reliable in area measurement due to spatial resolution constraints and the increasing zero drift over time (Figure 4B). While the proposed tracking system is free of those issues mentioned above, limitations still exist. In particular, the attribute of contact force and pressure are unavailable although they contribute to contact interactions (Essick et al., 2010; Huang et al., 2020; Teyssier et al., 2020; Xu et al., 2020). Due to the constraint of recording frequency, fast movements might fail in tracking since the hand image could be blurred. Meanwhile, the forearm needs to be recorded parallel with the y-axis of the color image coordinate. In so doing, the spatial hand velocity can be decomposed into the three orthogonal directions without additional markers to define the arm coordinate.

## 5.3 Further Applications in Human-to-Human Touch Interaction

Human touch each other with different intentions and a wide range of emotional states. In the classic theory of emotion, three dimensions of valence, arousal, and dominance, are typically employed for emotion assessments (Russell and Mehrabian, 1977; Russell, 1980). Indeed, using machine-controlled brush stimuli, the valence rating was reported to be tuned by the tangential stroking velocity (Löken et al., 2009; Essick et al., 2010; Ackerley et al., 2014a, 2014b; Croy et al., 2021). In the scenario of naturalistic human touch, our measurements could further facilitate the quantitative analysis regarding other correlates between contact attributes and the three emotional dimensions.

From the perspective of neurophysiology, changes in the skin's mechanics caused by physical contact could elicit different responses of peripheral afferents (Johnson, 2001; Yao and Wang, 2019; Xu et al., 2021a). For example, the firing frequency of C-tactile afferents is associated with the stroking velocity in an inverted-U shape relationship (Löken et al., 2009; Ackerley et al., 2014a; Liljencrantz and Olausson, 2014). Other Aβ afferents are suggested to support the identification of distinct emotional messages delivered by touch (Hauser et al., 2019b). Moving forward into this direction, measurements of naturalistic human contact can aid in uncovering how exactly afferents respond to such contact and contribute to different emotional percepts.

Affective touch is also believed to impact physiological arousal such as blood pressure, heart rate, respiration, ECG, EEG, and hormone level (Gallace and Spence, 2010; Sefidgar et al., 2016). Especially for infants, touch delivered by caregivers contributes to their social, cognitive, and physical development (Hertenstein, 2002; Van Puyvelde et al., 2019), where the underlying contact details would be meaningful to quantify. Additionally, many physical therapies, such as massage, rely on specific manipulation of the muscle and tissue of patients delivered by professional therapists. Those therapies create health benefits including relieving stress and pain, promoting blood circulation, and boosting mental wellness (Moyer et al., 2004). While the underlying mechanism is waiting to be further explored with the aid of physical skin contact tracking.

## 6 Conflict of Interest

The authors declare that the research was conducted in the absence of any commercial or financial relationships that could be construed as a potential conflict of interest.

## 7 Author Contributions

SX, CX, SM, HO, and GJG conceptualized and designed the study. SX and GJG developed the tracking system. SX, CX, and GJG performed the experiments. SX, CX, SM, and GJG analyzed and interpreted experimental results. SX, CX, and GJG drafted the manuscript. All authors edited and approved the manuscript.

## 8 Funding

This work is supported in part by grants from the National Science Foundation (IIS-1908115) and the National Institutes of Health (NINDS R01NS105241) to GJG. The funders had no role in study design, data collection and analysis, decision to publish, or preparation of the manuscript.

## 9 Acknowledgments

We would like to thank all the participants who participated in the experiments.

## 10 Reference

Ackerley, R., Backlund Wasling, H., Liljencrantz, J., Olausson, H., Johnson, R. D., and Wessberg, J. (2014a). Human C-tactile afferents are tuned to the temperature of a skin-stroking caress. *J. Neurosci.* 34, 2879–2883. doi:10.1523/JNEUROSCI.2847-13.2014.

Ackerley, R., Carlsson, I., Wester, H., Olausson, H., and Backlund Wasling, H. (2014b). Touch perceptions across skin sites: Differences between sensitivity, direction discrimination and pleasantness. *Front. Behav. Neurosci.* 8, 54. doi:10.3389/FNBEH.2014.00054.

Andreasson, R., Alenljung, B., Billing, E., and Lowe, R. (2018). Affective Touch in Human–Robot Interaction: Conveying Emotion to the Nao Robot. *Int. J. Soc. Robot.* 10, 473–491. doi:10.1007/s12369-017-0446-3.

Bucci, P., Cang, X. L., Valair, A., Marino, D., Tseng, L., Jung, M., et al. (2017). "Sketching CuddleBits: Coupled Prototyping of Body and Behaviour for an Affective Robot Pet," in *Proceedings of the 2017 CHI Conference on Human Factors in Computing Systems*, 3681–3692. doi:10.1145/3025453.3025774.

Cang, X. L., Bucci, P., Strang, A., Allen, J., Maclean, K., and Liu, H. Y. S. (2015). Different strokes and different folks: Economical dynamic surface sensing and affect-related touch recognition. in *ICMI 2015 - Proceedings of the 2015 ACM International Conference on Multimodal Interaction*, 147–154. doi:10.1145/2818346.2820756.

Cascio, C. J., Moore, D., and McGlone, F. (2019). Social touch and human development. *Dev. Cogn. Neurosci.* 35, 5–11. doi:10.1016/j.dcn.2018.04.009.

Christ, M., Braun, N., Neuffer, J., and Kempa-Liehr, A. W. (2018). Time Series FeatuRe Extraction on basis of Scalable Hypothesis tests (tsfresh – A Python package). *Neurocomputing* 307, 72–77. doi:10.1016/j.neucom.2018.03.067.

Coan, J. A., Schaefer, H. S., and Davidson, R. J. (2006). Lending a hand: Social regulation of the neural response to threat. *Psychol. Sci.* 17, 1032–1039. doi:10.1111/j.1467-9280.2006.01832.x.

Croy, I., Bierling, A., Sailer, U., and Ackerley, R. (2021). Individual Variability of Pleasantness Ratings to Stroking Touch Over Different Velocities. *Neuroscience* 464, 33–43. doi:10.1016/J.NEUROSCIENCE.2020.03.030.

Essick, G. K., McGlone, F., Dancer, C., Fabricant, D., Ragin, Y., Phillips, N., et al. (2010). Quantitative assessment of pleasant touch. *Neurosci. Biobehav. Rev.* 34, 192–203. doi:10.1016/J.NEUBIOREV.2009.02.003.

Gallace, A., and Spence, C. (2010). The science of interpersonal touch: An overview. *Neurosci. Biobehav. Rev.* 34, 246–259. doi:10.1016/j.neubiorev.2008.10.004.

Hauser, S. C., McIntyre, S., Israr, A., Olausson, H., and Gerling, G. J. (2019a). Uncovering Human-to-Human Physical Interactions that Underlie Emotional and Affective Touch Communication. in *2019 IEEE World Haptics Conference (WHC)*, 407–412. doi:10.1109/WHC.2019.8816169.

Hauser, S. C., Nagi, S. S., McIntyre, S., Israr, A., Olausson, H., and Gerling, G. J. (2019b). From Human-to-Human Touch to Peripheral Nerve Responses. in *2019 IEEE World Haptics Conference (WHC)* (IEEE), 592–597. doi:10.1109/WHC.2019.8816113.

Hertenstein, M. J. (2002). Touch: Its communicative functions in infancy. *Hum. Dev.* 45, 70–94.

Hertenstein, M. J., Holmes, R., Mccullough, M., and Keltner, D. (2009). The Communication of Emotion via Touch. *Emotion* 9, 566–573. doi:10.1037/a0016108.

Hertenstein, M. J., Keltner, D., App, B., Bulleit, B. A., and Jaskolka, A. R. (2006). Touch communicates distinct emotions. *Emotion* 6, 528–533. doi:10.1037/1528-3542.6.3.528.

Huang, C., Wang, Q., Zhao, M., Chen, C., Pan, S., and Yuan, M. (2020). Tactile Perception Technologies and Their Applications in Minimally Invasive Surgery: A Review. *Front. Physiol.* 11, 1601. doi:10.3389/FPHYS.2020.611596.

Johnson, K. O. (2001). The roles and functions of cutaneous mechanoreceptors. *Curr. Opin. Neurobiol.* 11, 455–461. doi:10.1016/S0959-4388(00)00234-8.

Jung, M. M., Cang, X. L., Poel, M., and Maclean, K. E. (2015). Touch challenge'15: Recognizing social touch gestures. in *Proceedings of the 2015 ACM International Conference on Multimodal Interaction (ICMI)* (New York, NY, USA), 387–390. doi:10.1145/2818346.2829993.

Liljencrantz, J., and Olausson, H. (2014). Tactile C fibers and their contributions to pleasant sensations and to tactile allodynia. *Front. Behav. Neurosci.* 8, 37. doi:10.3389/FNBEH.2014.00037.

Lo, C., Chu, S. T., Penney, T. B., and Schirmer, A. (2021). 3D Hand-Motion Tracking and Bottom-Up Classification Sheds Light on the Physical Properties of Gentle Stroking. *Neuroscience* 464, 90–104. doi:https://doi.org/10.1016/j.neuroscience.2020.09.037.

Löken, L. S., Wessberg, J., Morrison, I., McGlone, F., and Olausson, H. (2009). Coding of pleasant touch by unmyelinated afferents in humans. *Nat. Neurosci.* 12, 547–548. doi:10.1038/nn.2312.

Löning, M., Bagnall, A., Ganesh, S., Kazakov, V., Lines, J., and Király, F. J. (2019). sktime: A Unified Interface for Machine Learning with Time Series. *arXiv Prepr. arXiv1909.07872*.

McIntyre, S., Hauser, S. C., Kuztor, A., Boehme, R., Moungou, A., Isager, P. M., et al. (2021). The language of social touch is intuitive and quantifiable. *Psychol. Sci.*

Moyer, C. A., Rounds, J., and Hannum, J. W. (2004). A Meta-Analysis of Massage Therapy Research. *Psychol. Bull.* 130, 3–18. doi:10.1037/0033-2909.130.1.3.

Nath, T., Mathis, A., Chen, A. C., Patel, A., Bethge, M., and Mathis, M. W. (2019). Using DeepLabCut for 3D markerless pose estimation across species and behaviors. *Nat. Protoc.* 14, 2152–2176. doi:10.1038/s41596-019-0176-0.

Peters, R. M., Hackeman, E., and Goldreich, D. (2009). Diminutive Digits Discern Delicate Details: Fingertip Size and the Sex Difference in Tactile Spatial Acuity. *J. Neurosci.* 29, 15756–15761. doi:10.1523/JNEUROSCI.3684-09.2009.

Rezaei, M., Nagi, S. S., Xu, C., McIntyre, S., Olausson, H., and Gerling, G. J. (2021). Thin Films on the Skin, but not Frictional Agents, Attenuate the Percept of Pleasantness to Brushed Stimuli. in *2021 IEEE World Haptics Conference (WHC)*, 49–54. doi:10.1109/WHC49131.2021.9517259.

Romero, J., Tzionas, D., and Black, M. J. (2017). Embodied hands: modeling and capturing hands and bodies together. *ACM Trans. Graph.* 36. doi:10.1145/3130800.3130883.

Russell, J. A. (1980). A circumplex model of affect. *J. Pers. Soc. Psychol.* 39, 1161–1178.

Russell, J. A., and Mehrabian, A. (1977). Evidence for a three-factor theory of emotions. *J. Res. Pers.* 11, 273–294. doi:10.1016/0092-6566(77)90037-X.

Russo, V., Ottaviani, C., and Spitoni, G. F. (2020). Affective touch: A meta-analysis on sex differences. *Neurosci. Biobehav. Rev.* 108, 445–452. doi:10.1016/j.neubiorev.2019.09.037.

Rusu, R. B., and Cousins, S. (2011). 3D is here: Point Cloud Library (PCL). *Proc. IEEE Int. Conf. Robot. Autom.* doi:10.1109/ICRA.2011.5980567.

Sefidgar, Y. S., MacLean, K. E., Yohanan, S., Van Der Loos, H. F. M. H., Croft, E. A., and Garland, E. J. (2016). Design and Evaluation of a Touch-Centered Calming Interaction with a Social Robot. *IEEE Trans. Affect. Comput.* 7, 108–121. doi:10.1109/TAFFC.2015.2457893.

Silvera-Tawil, D., Rye, D., and Velonaki, M. (2014). Interpretation of Social Touch on an Artificial Arm Covered with an EIT-based Sensitive Skin. *Int. J. Soc. Robot.* 6, 489–505. doi:10.1007/s12369-013-0223-x.

Suresh, A. K., Goodman, J. M., Okorokova, E. V., Kaufman, M., Hatsopoulos, N. G., and Bensmaia, S. J. (2020). Neural population dynamics in motor cortex are different for reach and grasp. *Elife* 9, 1–16. doi:10.7554/ELIFE.58848.

Suvilehto, J. T., Nummenmaa, L., Harada, T., Dunbar, R. I. M., Hari, R., Turner, R., et al. (2019). Cross-cultural similarity in relationship-specific social touching. *Proc. R. Soc. B Biol. Sci.* 286, 20190467. doi:10.1098/rspb.2019.0467.

Taylor, J., Bordeaux, L., Cashman, T., Corish, B., Keskin, C., Sharp, T., et al. (2016). Efficient and precise interactive hand tracking through joint, continuous optimization of pose and correspondences. in *ACM Transactions on Graphics*, 1–12. doi:10.1145/2897824.2925965.

Teyssier, M., Bailly, G., Pelachaud, C., and Lecolinet, E. (2020). Conveying Emotions Through Device-Initiated Touch. *IEEE Trans. Affect. Comput.*, 1–1. doi:10.1109/TAFFC.2020.3008693.

Thompson, E. H., and Hampton, J. A. (2011). The effect of relationship status on communicating emotions through touch. *Cogn. Emot.* 25, 295–306. doi:10.1080/02699931.2010.492957.

Tsalamlal, M. Y., Ouarti, N., Martin, J.-C., and Ammi, M. (2014). Haptic communication of dimensions of emotions using air jet based tactile stimulation. *J. Multimodal User Interfaces 2014 91* 9, 69–77. doi:10.1007/S12193-014-0162-3.

Vallbo, Å., Löken, L., and Wessberg, J. (2016). "Sensual touch: A slow touch system revealed with microneurography," in *Affective Touch and the Neurophysiology of CT Afferents*, 1–30. doi:10.1007/978-1-4939-6418-5_1.

Van Puyvelde, M., Collette, L., Gorissen, A. S., Pattyn, N., and McGlone, F. (2019). Infants autonomic cardio-respiratory responses to nurturing stroking touch delivered by the mother or the father. *Front. Physiol.* 10, 1117. doi:10.3389/FPHYS.2019.01117.

Xu, C., He, H., Hauser, S. C., and Gerling, G. J. (2020). Tactile Exploration Strategies with Natural Compliant Objects Elicit Virtual Stiffness Cues. *IEEE Trans. Haptics* 13, 4–10. doi:10.1109/TOH.2019.2959767.

Xu, C., Wang, Y., and Gerling, G. J. (2021a). An elasticity-curvature illusion decouples cutaneous and proprioceptive cues in active exploration of soft objects. *PLoS Comput. Biol.* 17, e1008848. doi:10.1371/JOURNAL.PCBI.1008848.

Xu, C., Wang, Y., and Gerling, G. J. (2021b). Individual Performance in Compliance Discrimination is Constrained by Skin Mechanics but Improved under Active Control. *2021 IEEE World Haptics Conf. WHC 2021*, 445–450. doi:10.1109/WHC49131.2021.9517269.

Yao, M., and Wang, R. (2019). Neurodynamic analysis of Merkel cell–neurite complex transduction mechanism during tactile sensing. *Cogn. Neurodyn.* 13, 293–302. doi:10.1007/S11571-018-9507-Z/FIGURES/8.

Yohanan, S., and MacLean, K. E. (2012). The Role of Affective Touch in Human-Robot Interaction: Human Intent and Expectations in Touching the Haptic Creature. *Int. J. Soc. Robot.* 4, 163–180. doi:10.1007/s12369-011-0126-7.

Zheng, X., Shiomi, M., Minato, T., and Ishiguro, H. (2020). What Kinds of Robot's Touch Will Match Expressed Emotions? *IEEE Robot. Autom. Lett.* 5, 127–134. doi:10.1109/LRA.2019.2947010.

Zhou, Y., Habermann, M., Xu, W., Habibie, I., Theobalt, C., and Xu, F. (2020). Monocular Real-time Hand Shape and Motion Capture using Multi-modal Data. in *Proceedings of the IEEE/CVF Conference on Computer Vision and Pattern Recognition (CVPR)*, 5346–5355.